\documentclass[11pt,twoside]{article}
\usepackage{newpasp}
\usepackage{epsfig}

\index{Barnes, J. E.}

\begin{document}

\title{Gas Dynamics in Galaxy Mergers}

\author{J. E. Barnes}
\affil{Institute for Astronomy, University of Hawaii,
       2680 Woodlawn Drive, Honolulu HI, 96822}

\begin{abstract}
In interacting and merging galaxies, gas is subject to direct
hydrodynamic effects as well as tidal forces. One consequence of
interactions is the rapid inflows of gas which may fuel starbursts and
AGN. But gas dynamics is not limited to inflows; a small survey of
equal-mass and unequal-mass encounters produces a wide variety of
features, including plumes between galaxies, extended disks formed by
infall of tidal debris, and counterrotating nuclear disks.  An even
richer spectrum of behavior awaits better thermodynamic models for gas
in merging galaxies.
\end{abstract}


\section{Background}

It has long been recognized that interstellar material plays an
important role in galaxy interactions.  Spitzer \& Baade (1951)
described fast collisions of galaxies as essentially hydrodynamic
affairs.  While emphasizing that stellar dynamics governs the outcome
of galaxy encounters, Toomre \& Toomre (1972) anticipated more than a
decade of numerical work in suggesting that interactions `bring {\it
deep\/} into a galaxy a fairly {\it sudden\/} supply of fresh fuel'.

The mechanics of this fueling process seem fairly clear.  Negroponte
\& White (1983) found central concentrations of gas in simulated
mergers of gas-rich disk galaxies, but the low resolution of their
models left the physics uncertain.  Noguchi (1987, 1988) and Combes,
Dupraz, \& Gerin (1990) showed that tidal encounters trigger bar
formation in stellar disks and that {\it gravitational\/} torques
exerted by these bars extract angular momentum from the gas.
Hernquist (1989) reported rapid gas inflows in minor mergers where a
disk galaxy swallows a small companion; in these experiments the
gravitational torques driving the gas inward were generated by both
the stellar disk and the companion.  Barnes \& Hernquist (1991, 1996)
modeled major mergers between disk galaxies.  This work extended the
gravitational inflow picture to show how orbital decay could deliver
$\sim 5 \times 10^9 {\rm\,M_\odot}$ of gas within the central $100
{\rm\,pc}$ of a merger remnant; such massive gas clouds are common in
the central regions of infrared-selected galaxies, many of which are
merger remnants (Sanders \& Mirabel 1996).  Hernquist \& Barnes (1991)
simulated the formation of a counterrotating gas disk in the aftermath
of a major merger.  Misaligned or counterrotating disks are found in
both merger remnants (Schweizer 1982) and elliptical galaxies
(e.g.~Bender 1990), strengthening the hypotheses that merging disk
galaxies form elliptical galaxies (Toomre \& Toomre 1972) and that the
centralized starbursts which power luminous infrared galaxies
represent the formation of the cores of elliptical galaxies
(e.g.~Kormendy \& Sanders 1992).

Existing simulations are still incomplete in one key respect: they
don't really capture the {\it thermo\/}dynamics of the gas.  Early
studies used `sticky particles' to mimic effects of dissipation.
Hernquist and co-workers adopted `smoothed particle hydrodynamics'
(SPH) techniques which can include radiative heating and cooling.  But
in practice, radiative cooling is usually cut off below $10^4
{\rm\,K}$ to prevent catastrophic instabilities, and most of the gas
remains near this cut-off temperature throughout the simulation.
Consequently, simulations including radiative processes are all but
indistinguishable from those using an isothermal equation of state
(Barnes \& Hernquist 1996).

At this point there are two options.  One is to adopt the isothermal
gas model and survey a range of encounters.  The other is to try to
develop models which represent the thermodynamics of a multi-phase
ISM.  Below I discuss both.

\section{Isothermal Encounter Survey}

Isothermal simulations are relatively cheap, so an encounter survey is
possible with modest computers.  I chose four encounter geometries,
and ran each once with a galactic mass ratio of 1:1 and once with a
mass ratio of 3:1.  Table~1 lists inclinations $i$ and pericentric
arguments $\omega$ for the four encounter geometries; in the 3:1
encounters $i_1$ and $\omega_1$ refer to the more massive galaxy.
These experiments used the same bulge/disk/halo galaxy models and
close parabolic orbits as Barnes (1998, Ch.~4), but here each disk
included a gaseous component amounting to $12.5$\% of the disk mass.
In simulation units with $G = 1$, each galaxy in the 1:1 encounters,
and each large galaxy in the 3:1 encounters, has total mass $M_{\rm
bulge} \! + \! M_{\rm disk} \! + \! M_{\rm halo} = \frac{1}{16} \! +
\! \frac{3}{16} \! + \! 1 = 1.25$, half-mass radius $r_{\rm half}
\simeq 0.28$, rotation period $t(r_{\rm half}) \simeq 1.2$, and
binding energy $E = -1.07$; the gas has specific internal energy
$u_{\rm int} = 0.014$.  The simulations, each using a total of $N_{\rm
gas} \! + \! N_{\rm stars} \!  + \!  N_{\rm halo} = 24576 \! + \!
29696 \! + \! 32768 = 87040$ particles, were run with a new N-body/SPH
code featuring adaptive smoothing and time-stepping.

\begin{table}
\caption{Disk angles for survey}
\begin{tabular}{lrrrr}
\tableline
Geometry   & $i_1$ & $\omega_1$ & $i_2$ & $\omega_2$ \\
\tableline
DIRect     &     0 &          0 &    71 &         30 \\
RETrograde &   180 &          0 &  -109 &         30 \\
POLar      &    71 &         90 &  -109 &         90 \\
INClined   &    71 &        -30 &  -109 &        -30 \\
\tableline
\tableline
\end{tabular}
\end{table}

\begin{figure}
\begin{center}
\epsfig{figure=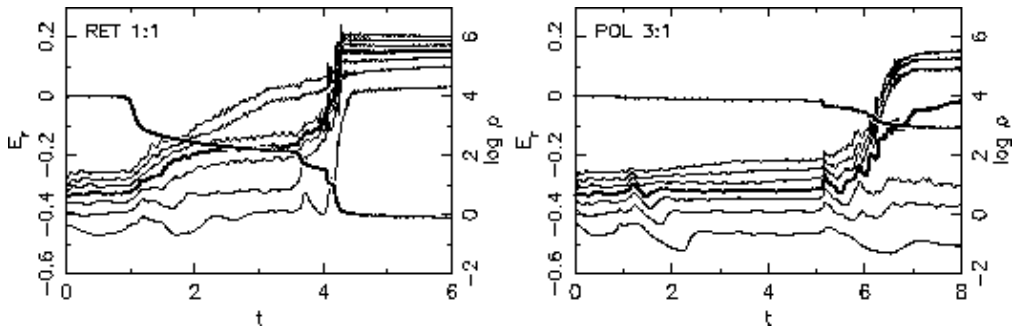,width=\textwidth}
\end{center}
\vspace{-0.4cm}
\caption{\small Evolution of two encounters.  Single falling curves
show $E_{\rm r}$, the energy lost to dissipation; rising curves show
the first through seventh octiles of the gas density, $\rho$.}
\label{fig01}
\end{figure}

Figure~\ref{fig01} summarizes the evolution of two of these
encounters.  Here $E_{\rm r}$ is the energy lost in radiative shocks,
while $\rho$ is the gas density.  Encounter RET 1:1 (left) is the most
dissipative of the eight studied here; radiative losses throughout
this calculation amount to $\sim 20$\% of its initial binding energy.
These losses occur in large-scale shocks as the two galaxies plow into
each other at $t = 1$, separate, and fall back together.  Gas
densities increase with each burst of dissipation, and by the end of
the simulation some $95$\% of the gas lies in a barely-resolved disk
at the center of the merger remnant.  In contrast, encounter POL 3:1
(right) is the least dissipative, losing only $\sim 5$\% of its
initial binding energy.  The first passage at $t = 1$, while close
enough to produce definite tidal features, barely registers in the
traces of $E_{\rm r}$ and $\rho$.  Later passages are more dramatic,
eventually driving about $60$\% into the central regions; most of the
remaining gas settles into a warped disk.

\begin{figure}
\begin{center}
\epsfig{figure=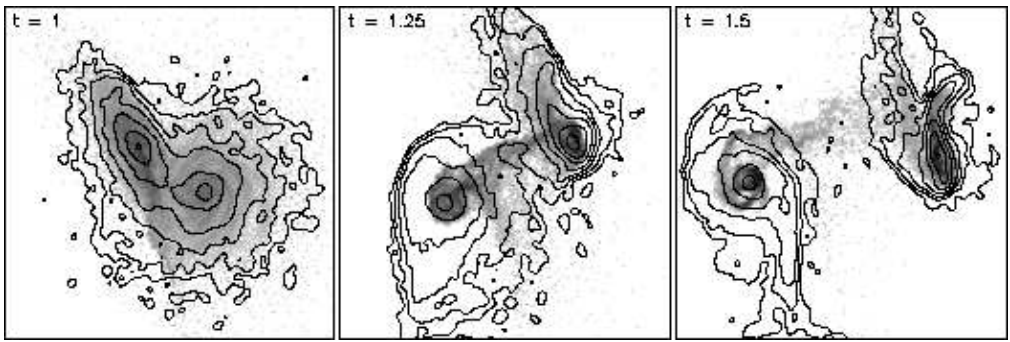,width=\textwidth}
\end{center}
\vspace{-0.4cm}
\caption{\small First passage of RET 1:1, viewed face-on to the
orbital plane.  Contours show stellar surface density in steps of one
magnitude; half-tones show gas.  Time appears in upper left of each
frame.  The first two frames are $1.25 \times 1.25$ length units; the
last frame is $1.5 \times 1.5$ length units.}
\label{fig02}
\end{figure}

The first passage of encounter RET 1:1, shown in Figure~\ref{fig02},
exemplifies some consequences of a violent {\it hydrodynamic\/}
interaction.  This encounter's geometry insures that most of the gas
suffers strong shocks as the galaxies intersect.  By the middle frame
shown here, much of the gas in the in-plane disk has been swept into
the center.  This inflow is driven not by gravitational torques but by
hydrodynamic forces; the gas loses its spin angular momentum by
colliding with gas in its companion.  Gas which escapes being swept
inward forms a plume connecting the two galaxies; such structures may
be fairly common in interpenetrating encounters (e.g.~Condon et
al.~1993).

\begin{figure}
\begin{center}
\epsfig{figure=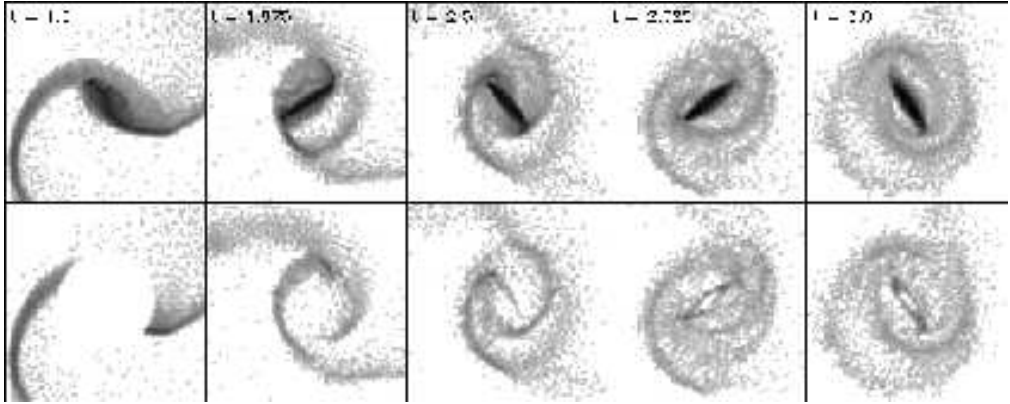,width=\textwidth}
\end{center}
\vspace{-0.4cm}
\caption{\small Evolution of large disk in DIR 3:1 after first
passage.  The upper frames shows all gas; the lower frames show gas
re-accreted from tidal features.  All frames are $0.8 \times 0.8$
length units.}
\label{fig03}
\end{figure}

After a tidal encounter, bridges and tails develop from any disk which
suffers a reasonably direct passage.  While the ends of such features
may escape, much of this material is bound to fall back onto its
parent -- or, if the galaxies have already merged, onto their combined
hulks.  An example of such re-accretion is illustrated in
Figure~\ref{fig03}.  Here the upper row shows gas in the larger disk
of encounter DIR 3:1 responding after the direct passage of its
lighter companion; note the pronounced bar typically formed in such
passages.  The lower row shows only the gas which has fallen back from
the tidal features.  Returning on elongated trajectories, this
material is forced onto more circular orbits by shocks, some of which
are visible as narrow curvilinear features.  By the last frame the
re-accreted gas has built up a large disk surrounding the bar.

\begin{figure}
\begin{center}
\epsfig{figure=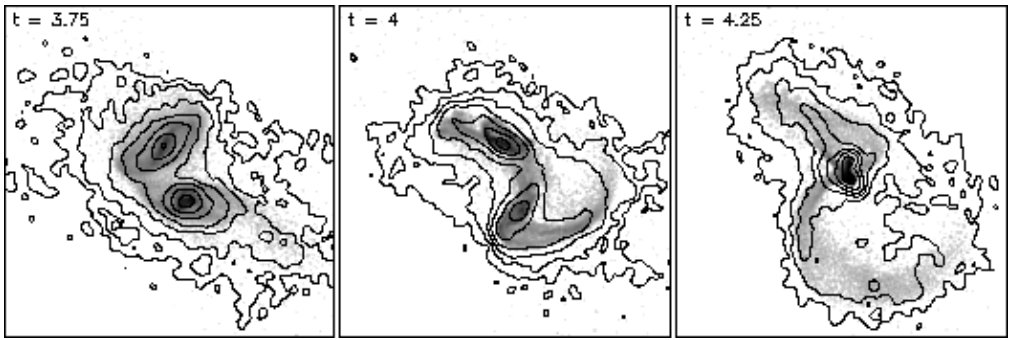,width=\textwidth}
\end{center}
\vspace{-0.4cm}
\caption{\small Second passage of POL 1:1.  Contours show stellar
surface density; half-tones show gas.  The first two frames are $1.25
\times 1.25$ length units; the last frame is $1.5 \times 1.5$ length
units.}
\label{fig04}
\end{figure}

One consequence of this disk rebuilding is that the gas can be highly
sensitive to tidal perturbations on subsequent passages.  Stellar
disks can't cool down once tidally heated, so they are less responsive
and generate broader structures.  Figure~\ref{fig04} illustrates the
different responses of gas and stars in the second passage of
encounter POL 1:1.  The first frame shows two fairly relaxed gas
disks, each wreathed in stellar debris; at this point the stars and
gas have roughly similar distributions.  In the second frame the disks
have moved past each other, and a pronounced tail extends to the right
of the lower galaxy.  By the third frame the gas tail extends far
beyond the stellar tail, crossing several contours of the stellar
distribution; this frame also shows the two galaxies in the process of
merging.

In these eight encounters, between $50$\% and $95$\% of the gas falls
into a compact cloud at the center of each merger remnant.  This is
consistent with the results of earlier calculations (Negroponte \&
White 1983; Noguchi 1988; Hernquist 1989; Barnes \& Hernquist 1991).

\begin{figure}
\begin{center}
\epsfig{figure=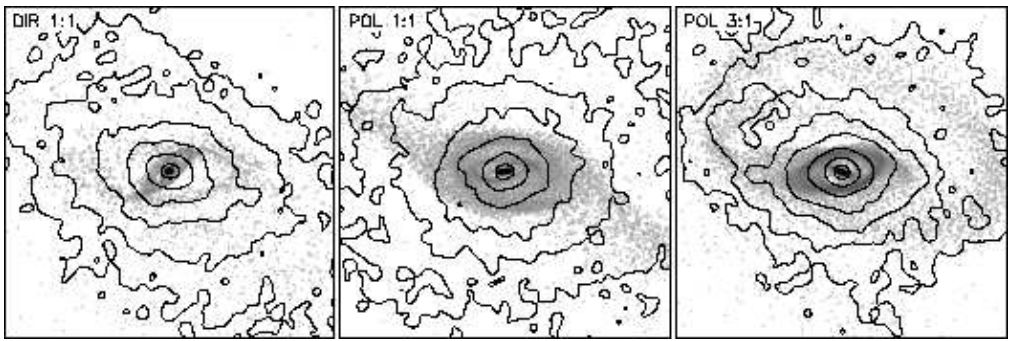,width=\textwidth}
\end{center}
\vspace{-0.4cm}
\caption{\small Gas disks in remnants of encounters DIR 1:1 (left),
POL 1:1 (middle), and POL 3:1 (right).  Contours show stellar surface
density; half-tones show gas.  All are viewed $30^\circ$ from edge-on
to the stellar distribution.  The first two frames are $0.5 \times
0.5$ length units; the third frame is $1.0 \times 1.0$ length units.}
\label{fig05}
\end{figure}

Most of the gas which doesn't wind up in the central cloud nonetheless
falls back into the remnant and settles onto closed, non-intersecting
orbits.  Thus at later times these remnants develop extra-nuclear
disks and rings of gas with rather complex morphologies, as shown by
the three examples in Figure~\ref{fig05}.  The remnant of encounter
DIR 1:1 contains {\it two\/} rings, both rather sparse; the outer one
lies within $\sim 10^\circ$ the equatorial plane, while the inner one
is tilted by $\sim 70^\circ$.  The origin of this tilt is unclear; the
tilted ring is largely made up of gas from the $i_2 = 71^\circ$ disk,
but it's born in a plane roughly perpendicular to the initial plane of
its progenitor.  In the remnant of POL 1:1 the extra-nuclear disk
contains $\sim 30$\% of the gas and exhibits an integral-sign warp;
this remnant also has a compact nuclear disk, to be discussed below.
Finally, the remnant of POL 3:1 has several nested disks and an
extended off-center ring composed of material recently accreted from
tidal debris.  While the outer contour of the stellar distribution
roughly parallels the gas ring, the stellar density falls
monotonically with distance from the center; unlike the gas, the stars
don't form a true ring.

\begin{figure}
\begin{center}
\epsfig{figure=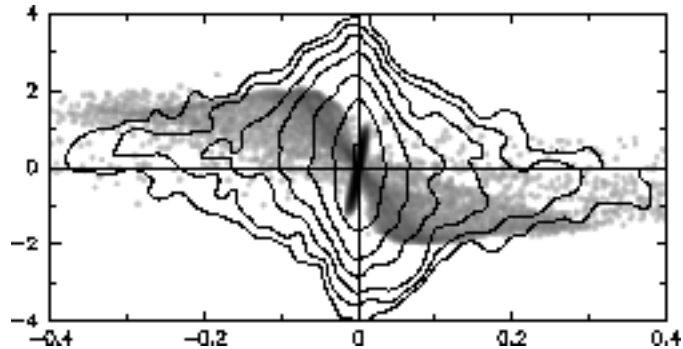,width=3.5in}
\end{center}
\vspace{-0.4cm}
\caption{\small Line-of-sight velocities in remnant POL 1:1, as viewed
edge-on to the equatorial plane.  Contours show the stellar
distribution within a slit along the major axis; half-tone shows {\it
all\/} gas.}
\label{fig06}
\end{figure}

Some of the remnants in this survey have {\it counter\/}rotating
nuclear disks.  The most striking example was produced by encounter
POL 1:1; Figure~\ref{fig06} shows that the outer gas disk and the
stellar component both rotate in one direction, while the inner gas
disk rotates in the other direction.  This encounter had the same
initial disk angles $i_1$, $\omega_1$, $i_2$, and $\omega_2$ as did a
previous example of counterrotation (Hernquist \& Barnes 1991), but
here the first passage was considerably closer.  Counterrotation seems
insensitive to details of the encounter and simulation code, though
the physical mechanism involved is not obvious from a cursory
inspection of the simulations.  More work is needed to determine the
roles of gravitational and hydrodynamic forces in creating
counterrotating disks and the {\it range\/} of disk angles yielding
outcomes like the one in Figure~\ref{fig06}.

\section{Towards Proper Thermodynamic Models}

The forgoing simulations presume that a good part of the ISM behaves
somewhat like an isothermal gas at a temperature of $10^4 {\rm\,K}$.
This is probably not a {\it desperately\/} bad approximation on scales
of order a kpc, but it doesn't address some of the most interesting
gas-dynamical effects in merging galaxies -- for example, starbursts
(Larson \& Tinsley 1978), infrared emission (Joseph \& Wright 1985),
molecular content (Young et al. 1984), and superwinds (Heckman et
al. 1996).  All but the last of these involve gas which has cooled to
well below $10^4 {\rm\,K}$.

It's hard to model the cold component with any degree of realism;
individual GMCs are about $10^{-4}$ times the total mass of the ISM in
a galaxy like ours, and key processes are incompletely understood.
One brave attempt by Gerritsen \& Icke (1997) allowed gas to cool
below $10^4 {\rm\,K}$, and defined sites of star formation via a
simple Jeans criterion; the radiation field of the stars, evaluated in
the optically thin limit, was used to heat the gas.  This yielded a
self-regulating medium with two phases at temperatures of $10^2
{\rm\,K}$ and $10^4 {\rm\,K}$; as a bonus, the simulations also
produced a Schmidt law with index $n \simeq 1.3$.  These
successes may inhere to almost any approach which balances radiative
cooling with heating by star formation.

In infrared-selected galaxies the ISM reprocesses $90$\% to $99$\% of
the total luminosity (Sanders \& Mirabel 1996), and this reprocessing
has thermodynamic consequences (e.g.~Maloney 1999) missing in
optically thin models.  Radiative reprocessing could be approximated
in SPH by a Monte-Carlo procedure or by solving the diffusion
equations, but galactic-scale models will require dramatic
improvements in both spatial and temporal resolution; moreover,
turbulent, magnetic, and radiation pressure terms may be needed as
these probably supplant thermal pressure in a starburst's ISM.  This
is a tall order for simulators!

If models of cold gas in starbursts are beyond the reach of current
simulations then star formation may be included phenomenologically;
for example, by setting $\tau_{\rm sf} \propto \rho^{-n}$, where
$\tau_{\rm sf}$ is the star formation timescale, $\rho$ is the local
gas density, and $n = 0.5$ for a Schmidt law (Katz 1992; Mihos \&
Hernquist 1994).  This approach seems to reproduce nuclear starbursts,
but doesn't readily explain {\it extended\/} star formation in systems
like NGC~4038/9 (Mirabel et al.~1998) and NGC~3690/IC~694.  A more
general approach could use the local dissipation rate $\dot{u}$ as a
second parameter; for example, setting $\tau_{\rm sf} \propto
\rho^{-n} \dot{u}^{-m}$.  This might approximate Jog \& Solomon's
(1992) model for triggered star formation via compression of molecular
clouds.

While cold gas is problematic, it's more straightforward to include
hot gas in simulations of galactic collisions.  If cooling is simply
turned off, gas can be shock-heated to $\sim 10^5 {\rm\,K}$ in the
early stages of an encounter, and finally to the virial temperature of
$\sim 10^6 {\rm\,K}$ in a merger (e.g.~Barnes 1998, Ch.~8).  But such
simulations fail to reproduce the superwinds often seen in starburst
galaxies; purely mechanical heating doesn't provide enough energy to
drive bulk outflows.  Injection of mass and energy by stellar winds,
supernovae, or AGN is probably required if the simulations are to
resemble real galaxies.

\section{Future Directions}

While stellar-dynamical modeling of galaxy collisions seems a fairly
mature business, the same isn't true of simulations including gas
dynamics.  Existing codes may do an acceptable job of simulating the
behavior of a smooth gas at $\sim 10^4 {\rm\,K}$, but the relationship
between this hypothetical stuff and the interstellar medium of real
galaxies is unclear.  I close by listing some possible directions for
further work:

\begin{itemize}

\item
True 3-D modeling of thin disks.  In published SPH calculations the
smoothing radius exceeds the vertical scale height of the gas.  An
order of magnitude more gas particles are needed to vertically resolve
thin disks.

\item
Interesting temperature structures.  As noted above, attempts to
include cold gas face many problems.  Hot gas is more easily
implemented, but source terms representing stellar processes are
probably required to match the observations of this component.

\item
Self-regulating star formation.  Until simulations including cold gas
become practical it may be difficult to include ``feedback'' from star
formation.  Nonetheless, it may be worth trying simple tricks; e.g.,
disallowing star formation in recently-shocked gas.

\item
Magnetic field diagnostics.  Polarization maps of continuum emission
can provide information on flows in galaxies (e.g.~Beck et al. 1999).
Simulations including magnetic fields may be able to predict such
polarization patterns.

\end{itemize}

\acknowledgements
Support for this work was provided by NASA grant NAG 5-8393 and by
Space Telescope Science Institute grant GO-06430.03-95A.

\vspace{0.5cm}

{\small

\noindent {\bf van der Hulst:\/} How long do the counterrotating disks
and warps in the models survive?

\noindent {\bf Barnes:\/} In DIR 1:1 the tilt of the inner ring
decreases with time, while in POL 1:1 the counterrotating nuclear disk
seems quite stable.  Warps should persist {\it at least\/} as long as
continued infall from tidal tails feeds in misaligned material.

\noindent {\bf Dwarakanath:\/} Is there an initial parameter space
which does not lead to counterrotating gas disks at the center?

\noindent {\bf Barnes:\/} Only one of the 1:1 remnants has a {\it
substantial\/} counter rotating disk.  And only one of the 3:1
remnants has a counterrotating disk of any kind.  Counterrotation
remains the exception rather than the rule.

\noindent {\bf Mathews:\/} To what extent could the extreme central
concentration of gas in post-merger galaxies be due to numerical
viscosity?

\noindent {\bf Barnes:\/} Barnes \& Hernquist (1996) varied the
artificial viscosity by a factor of three in either direction without
changing the resulting central concentration of the gas.  Moreover,
SPH and sticky particle codes give similar results.  It's unlikely
that numerical effects play a big role, but proving this beyond a
shadow of a doubt is not easy!

}


\begin{references}

{\small

\reference
Barnes, J.E. 1998, in Galaxies: Interactions and Induced Star
Formation, eds.~D. Friedli, L. Martinet, \& D. Pfenniger
(Springer-Verlag: Berlin), 275

\reference
Barnes, J.E. \& Hernquist, L.E. 1991, \apj, 370, L65

\reference
Barnes, J.E. \& Hernquist, L. 1996, \apj, 471, 115

\reference
Beck, R. et al. 1999, Nature, 397, 324

\reference
Bender, R. 1990, in Dynamics and Interactions of Galaxies,
ed.~R. Wielen (Springer, Berlin), 232

\reference
Combes, F., Dupraz, C., \& Gerin, M. 1990, in Dynamics and
Interactions of Galaxies, ed.~R. Wielen (Springer, Berlin), 205

\reference
Condon, J.J. et al. 1993, \aj, 105, 1730

\reference
Gerritsen, J.P.E. \& Icke, V. 1997, \aap, 325, 972

\reference
Heckman, T.M. et al. 1996, \apj, 457, 616

\reference
Hernquist, L. 1989, Nature, 340, 687

\reference
Hernquist, L. \& Barnes J.E. 1991, Nature, 354, 210

\reference
Jog, C.J. \& Solomon, P.M. 1992, \apj, 387, 152

\reference
Joseph, R.D. \& Wright, G.S. 1985, \mnras, 214, 87

\reference
Katz, N. 1992, \apj, 391, 502

\reference
Kormendy, J. \& Sanders, D. 1992, \apj, 390, L53

\reference
Larson, R.B. \& Tinsley, B.M. 1978, \apj, 219, 46

\reference
Maloney, P.R. 1999, \aaps, 266, 207

\reference
Mihos, J.C. \& Hernquist, L. 1994, \apj, 437, 611

\reference
Mirabel, I.F. et al. 1998, \aap, 333, L1

\reference
Negroponte, J. \& White, S.D.M. 1983, \mnras, 205, 1009

\reference
Noguchi, M. 1987, \mnras, 228, 635

\reference
Noguchi, M. 1988, \aap, 203, 259

\reference
Sanders, D.B. \& Mirabel, I.F. 1996, \araa, 34, 749

\reference
Schweizer, F. 1982, \apj, 252, 455

\reference
Spitzer, L. \& Baade, W. 1951, \apj, 113, 413

\reference
Toomre, A. \& Toomre, J. 1972, \apj, 405, 142

\reference
Young, J.S. et al. 1984, \apj, 287, L65

}
\end{references}
\end{document}